\documentclass{article}
\usepackage[a4paper, left=20mm, right=20mm, top=25mm, bottom=25mm]{geometry}
\usepackage{bm,amssymb,amsfonts,amsmath,graphicx}
\newcommand{\lbl}[1]{\label{#1}}
\newcommand{\bD}{{\mathbf{D}}}
\newcommand{\bmn}{{\bm{n}}}

\newcommand{\bmalpha}{{\bm{\alpha}}}
\newcommand{\bmtheta}{{\bm{\theta}}}

\newcommand{\bmA}{{\mathbf{A}}}

\begin{document}

\begin{center}
  
  \textbf{\Large From $\chi^2$ to Bayesian model comparison and Levy
    expansions of Bose-Einstein correlations in $e^+e^-$
    reactions\footnote{7th Workshop on Particle Correlations and
      Femtoscopy, September 20--24, 2011, Tokyo} }\\[12pt]

\textit{Michiel B.\ De Kock, Hans C.\ Eggers}, \\
  \textit{Department of Physics, University of Stellenbosch,
  ZA--7600 Stellenbosch, South Africa}\\[12pt]

\textit{Tam\'as Cs\"org\H{o}}\\
  \textit{Wigner RCP, RMKI, H--1525 Budapest 114, P.\ O.\ Box 49, Hungary}
\end{center}

\abstract{The usual $\chi^2$ method of fit quality assessment is a
  special case of the more general method of Bayesian model comparison
  which involves integrals of the likelihood and prior over all
  possible values of all parameters.  We introduce new
  parametrisations based on systematic expansions around the stretched
  exponential or Fourier-transformed L\'evy source distribution, and
  utilise the increased discriminating power of the Bayesian approach
  to evaluate the relative probability of these models to be true
  representations of a recently measured Bose-Einstein correlation
  data in $e^+e^-$ annihilations at LEP.}

\section{Bayes factors}

\noindent
The Bayesian definition of probability differs radically from the
conventional ``frequentist'' one, necessitating the overhaul of many
concepts and techniques used in statistics and its applications. Since
its introduction in 1900 \cite{Pea00a}, the $\chi^2$ statistic has
become the standard criterion for goodness of fit in physics and many
other disciplines, while Laplace's Bayesian approach \cite{Lap74a}
remained largely forgotten until revived by Jeffreys
\cite{Jef61a}. Later refinements such as the Maximum Likelihood occupy
a middle ground between the two approaches.

In this contribution, we demonstrate the use of one Bayesian technique
in the simple context of fitting or, more generally, the quantitative
assessment of evidence in favour of a hypothesis $H_1$ as a
description of given data, compared to a rival hypothesis $H_2$. We do
so by analysing the concrete example of binned data for the
correlation function $C_2(Q)$ in the four-momentum difference $Q =
\sqrt{-(p_1-p_2)^2}$ as published recently by the L3 Collaboration
\cite{L311a}.

Suppose we have data $\bD = \{Q_1,\ldots,Q_n\}$ consisting of $n$
measurements of particle four-momen\-tum differences, assumed to be
mutually independent as is customary in femtoscopy.  Typically, the
experimentalist will want to test how well various parametrisations
fit the data. For the purposes of Bayesian analysis, a given
parametrisation $y(Q\,|\,\bmtheta_m)$ with $N_m$ free parameters
$\bmtheta_m = \{\theta_{m1}, \theta_{m2}, \ldots, \theta_{mN_m}\}$ is
considered a ``model'' or ``hypothesis'' $H_m$. The starting point is
the \textit{odds in favour of model $H_m$ compared to a different
  model $H_\ell$''}, defined as the ratio
$p(H_m\,|\,\bD)/p(H_\ell\,|\,\bD)$, while the \textit{evidence for
  $H_m$ versus $H_\ell$} is the logarithm\footnote{We use $\lg =
  \log_2$; other base units can be substituted as preferred.} of the
odds.  Use of Bayes' Theorem for both hypotheses yields
\begin{align}
  \lbl{bqc}
  \frac{p(H_m\,|\,\bD)}{p(H_\ell\,|\,\bD)}
  &= \frac{p(\bD\,|\,H_m)\,p(H_m)}{\sum_k p(\bD\,|\,H_k)\,p(H_k)}
  \;  \frac{\sum_k p(\bD\,|\,H_k)\,p(H_k)}{p(\bD\,|\,H_\ell)\,P(H_\ell)}
  = \frac{p(\bD\,|\,H_m)}{p(\bD\,|\,H_\ell)} \cdot 
  \frac{p(H_m)}{p(H_\ell)}.
\end{align}
The evidence of $H_m$ versus $H_\ell$ is therefore the same as the
\textit{Bayes factor} $B_{m\ell} =\lg
[p(\bD\,|\,H_m)/p(\bD\,|\,H_\ell)]$ if there is no a priori reason
to prefer $H_m$ above $H_\ell$ and therefore $p(H_m) = p(H_\ell) =
1/2$.  A large Bayes factor says that the evidence for $H_m$ is
stronger than the evidence for $H_\ell$ and vice versa.  It can be
written as a ratio of integrals over the respective parameter spaces
of $\bmtheta_m$ and $\bmtheta_\ell$,
\begin{align} 
  \lbl{bqd}
  B_{m\ell}
  &= \lg \frac{p(\bD\,|\,H_m)}{p(\bD\,|\,H_\ell)}
  = \lg \frac{\int d\bmtheta_m\; p(\bD|\bmtheta_m,H_m)\,p(\bmtheta_m|H_m)}
          {\int d\bmtheta_\ell\; p(\bD|\bmtheta_\ell,H_\ell)\,p(\bmtheta_\ell|H_\ell)}.
\end{align}
Solving the high-dimensional integrals will often be an arduous
task. Fortunately, the independence of the measurements implies that
the likelihood $p(\bD\,|\,\bmtheta_m,H_m)$ factorises into the product
of likelihoods for individual data points, which by assumption have
the same form,
\begin{align}
  \lbl{bqe}
  p(\bD\,|\,\bmtheta_m,H_m) 
  = \prod_i p(Q_i\,|\,\bmtheta_m,H_m)
  \approx [p(Q\,|\,\bmtheta_m,H_m)]^n.
\end{align}
Due to the large exponent, even the slightest nonuniformity in
$p(Q\,|\,\bmtheta_m,H_m)$ will lead to the development of a strong
peak in parameter space for the overall likelihood, situated at the
maximum likelihood point $\hat\bmtheta_m$. An asymmetric prior
$p(\bmtheta_m\,|\,H_m)$ will shift the peak to a value $\bmtheta_m^*$,
but it will not materially affect the width of the peak or its
differentiability. Unless the shifted peak falls on a boundary of the
parameter space or happens to be nondifferentiable, it can therefore
be expanded around $\bmtheta_m^*$ \cite{Kas95a}:
\begin{align}
  \lbl{bqf}
  p(\bD\,|\,\bmtheta_m,H_m)\,p(\bmtheta_m\,|\,H_m)
  \simeq 
  p(\bD\,|\,\bmtheta^*,H_m)\,p(\bmtheta^*\,|\,H_m)
  \exp\left[ -\frac{1}{2} (\bmtheta_m-\bmtheta_m^*) 
    \bmA^{-1} (\bmtheta_m-\bmtheta_m^*) \right]
\end{align}
where $\bmA^{-1}$ is the Hessian of the expansion 
\begin{align}
  \lbl{bqg}
  A_{ij}^{-1} = -
  \frac{\partial^2 \ln [p(\bD\,|\,\bmtheta_m,H_m)\,p(\bmtheta_m\,|\,H_m)]}
  {\partial \theta_{mi} \,\partial\theta_{mj}}\biggr|_{\bmtheta_m^*}
\end{align}
and $\bmA$ is the parameter covariance matrix.  As more data is
accumulated, the peak narrows so that we can neglect the fact that
parameters may have finite ranges. Integrating the above as if it were
a Gaussian, one obtains Laplace's result \cite{Lap74a}
\begin{align}
  \lbl{bqh}
  \int_{-\infty}^{+\infty} d\bmtheta\;
  p(\bD\,|\,\bmtheta_m,H_m)\,p(\bmtheta_m\,|\,H_m)
  &\simeq 
  p(\bD\,|\,\bmtheta_m^*,H_m)\,p(\bmtheta_m^*\,|\,H_m)
  \sqrt{(2\pi)^{N_m} \det \bmA_m}\,,
\end{align}
which under the stated assumptions is a good approximation of the
full-blown integral appearing in Eq.~(\ref{bqd}) if $n \gtrsim 20
N_m$. The Bayes factor becomes simply the difference 
\begin{align}
  \lbl{bqi}
  B_{m\ell} &\simeq h_\ell - h_m \\
  \lbl{bqj}
  h_k &\equiv
  -\lg\left[ 
  p(\bD\,|\,\bmtheta_k^*,H_k)\,p(\bmtheta_k^*\,|\,H_k) \sqrt{(2\pi)^{N_k} \det \bmA_k}
    \right].
\end{align}
Evidence $h_k$ can be determined for any single model $H_k$, but has
no meaning on its own; only differences $h_\ell-h_m$ are meaningful in
quantifying the probability for $H_m$ to be true compared to $H_\ell$,
\begin{align}
  \lbl{bqk}
  \frac{p(H_m\,|\,\bD)}{p(H_\ell\,|\,\bD)}
  \simeq 2^{h_\ell-h_m}.
\end{align}

\section{Relationship to $\chi^2$ and the Maximum Likelihood}

\noindent
The Bayesian results obtained above differ from the traditional
Maximum Likelihood Estimate (MLE), which ignores the priors
$p(\bmtheta_m\,|\,H_m)$ and approximates the integral (\ref{bqd}) to
the maxima of the likelihoods,
\begin{align} 
  \lbl{cmc}
  B_{m\ell}
  &= \lg \frac{\int d\bmtheta_m\; p(\bD|\bmtheta_m,H_m)\,p(\bmtheta_m|H_m)}
          {\int d\bmtheta_\ell\; p(\bD|\bmtheta_\ell,H_\ell)\,p(\bmtheta_\ell|H_\ell)}
  \simeq \lg \frac{p(\bD|\hat\bmtheta_m,H_m)}
          {p(\bD|\hat\bmtheta_\ell,H_\ell)}.
\end{align}
The traditional $\chi^2$ goodness-of-fit is related to the above as
follows. The measurements $\{Q_i\}$ are binned into bins
$b=1,\ldots,B$ with bin midpoints $Q_b$, yielding the histogram
version of the data, $\bD = \{n_b\}_{b=1}^B$ with $\sum_b n_b=1$. The
most general ``parametrisation'' of the histogram contents is then the
multinomial with $\bmalpha = \{\alpha_b\}_{b=1}^B$ the set of
Bernoulli probabilities with $B-1$ degrees of freedom,
\begin{equation} 
  \lbl{cmd} 
  p(\bmn\,|\,\bmalpha,n) = n! \prod_{b=1}^B \frac{\alpha_b^{n_b}}{n_b!} \,,
\end{equation}
which on use of the Stirling approximation becomes, up to a
normalisation constant,
\begin{align} 
  \lbl{btf} 
  p(\bmn\,|\,\bmalpha,n)
  &= c\cdot \exp\left[ -\sum_b n_b \ln \frac{n_b}{n\alpha_b}\right].
\end{align}
Expanding the free parameters $\bmalpha$ around the measured data
$\bmn$ and truncating
\begin{align} 
  \lbl{cme} 
  p(\bmn\,|\,\bmalpha,n) 
  &= c\cdot 
  \exp\left[ -\sum_b \left( \frac{(n \alpha_b - n_b)^2}{2 n_b} 
    - \frac{(n\alpha_b - n_b)^3}{3 n_b^2}
    + \ldots\right) \right]
  \simeq c\cdot 
  \exp\left[ -\frac{1}{2}\sum_b \frac{(n \alpha_b - n_b)^2}{n_b} \right],
\end{align}
we can identify the multinomial quantities with the measured
correlation functions at mid-bin points $Q_b$ by setting\footnote{$I$
  is an arbitrary large integer to ensure that $I\,C_2(Q_b)$ is an
  integer. As it eventually cancels out, its size is immaterial.} $n_b
\to I\,C_2(Q_b)$, $C = \sum_b C_2(Q_b)$, and $n\to I\, C$. The $n_b$ in
the denominator is almost equal to the measured bin variances
$n_b\simeq \sigma^2(n_b) = I^2\,\sigma^2(C_2(Q_b))$ so that the
quadratic term is
\begin{align} 
  \lbl{cmf} 
  \frac{(n \alpha_b - n_b)^2}{2\,n_b} \simeq 
  \frac{[C_2(Q_b)-y(Q_b\,|\,\hat\bmtheta_m)]^2}{2\,\sigma(C_2(Q_b))^2},
\end{align}
where $n\alpha_b/I \to y(Q_b\,|\,\hat\bmtheta_m)$, which includes all
the constants, is the unnormalised parametrisation for $C_2(Q)$ in
common use. Comparing this to the usual definition
\begin{align} 
  \lbl{cmg} 
  \chi^2 = \sum_b \frac{[C_2(Q_b) - y(Q_b\,|\,\hat\bmtheta_m)]^2}{\sigma(C_2(Q_b))^2},
\end{align}
we see that the maximum likelihood is approximately equal to
\begin{align} 
  \lbl{cmh} 
  p(\bD|\hat\bmtheta_m,H_m) 
  \simeq e^{-\chi^2/2},
\end{align}
so that $\chi^2$ is seen to be an approximation of the Bayes
formulation, using only a single point in the parameter space
$\bmtheta_m^* \equiv \hat\bmtheta_m$ and thereby effectively assuming
a uniform prior. Furthermore, $\chi^2$ truncates the expansion of
(\ref{cme}); this is probably the approximation most vulnerable to
criticism.

\section{Parametrisations and L\'evy-based polynomial expansions}

\noindent
We now apply the above general ideas to the specific case of the
various parametrisations shown in Table 1 for the correlation function
data for two-jet events published by the L3 Collaboration
\cite{L311a}. Hypotheses $H_1$ to $H_3$ are taken from the L3 paper.
Realising that it is important to quantify the degree of deviation of
Bose-Einstein correlation data from the Gaussian or the exponential
shape, the L3 Collaboration also studied a ``Laguerre expansion'' as
well as the symmetric L\'evy source distribution, characterized by the
stretched-exponential correlation function of hypothesis $H_2$. In
$H_4$ and $H_5$, we propose a new expansion technique that measures
deviations from $H_2$ in terms of a series of ``L\'evy polynomials''
that are orthogonal to the characteristic function of symmetric L\'evy
distributions, generalising the results presented in
Ref.~\cite{Csorgo:2000pf}. 
\begin{align}
  \lbl{lpc}
  L_1(x\,|\,\alpha) = \det\left(\begin{array}{c@{\hspace*{8pt}}c}
      \mu_{0,\alpha} & \mu_{1,\alpha} \\ 
      1 & x \end{array} \right)
  \qquad
  L_2(x\,|\,\alpha) = \det\left(\begin{array}{c@{\hspace*{8pt}}c@{\hspace*{8pt}}c}
      \mu_{0,\alpha} & \mu_{1,\alpha} & \mu_{2,\alpha} \\ 
      \mu_{1,\alpha} & \mu_{2,\alpha} & \mu_{3,\alpha} \\ 
      1 & x & x^2 \end{array} \right) \qquad \text{etc.}
\end{align}
where $\mu_{r,\alpha} = \int_0^\infty dx\;x^{r} f(x\,|\,\alpha) =
\tfrac{1}{\alpha}\,\Gamma( \tfrac{r+1}{\alpha})$. These reduce, up to
a normalisation constant, to the Laguerre polynomials for
$\alpha=1$. Figure 1 displays two examples for various values of
$\alpha$. Polynomials cannot be both orthogonal and derivatives for
transcendental weight functions \cite{Erd53a}, and therefore in $H_6$
and $H_7$ we also investigated nonorthogonal derivative functions of
the stretched exponential\footnote{Note the absence of the
  $[1+\varepsilon Q]$ long-range correction term. L3 demonstrated that
  this term vanishes if the dip, the non-positive definiteness of
  $C_2(Q)-1$, is taken into account by the parametrisation elsewhere,
  e.g.\ by the cosine in $H_3 $ and by the first-order polynomials in
  $H_4$ and $H_5$, resulting in $\epsilon$ values consistent with
  zero.}.\\ %
\renewcommand{\arraystretch}{1.5}
\begin{center}
  \begin{tabular}{|rllc|}
    \hline & Hypothesis & Functional form & $N_m$\\ \hline
    $H_1$ & Gauss & $\displaystyle \gamma[1+\varepsilon Q]\,\left[1 + \lambda e^{-R^2 Q^2}\right]$ & 4 \\
    $H_2$ & Stretched Exponential & $\displaystyle \gamma[1+\varepsilon Q]\,\left[1 + \lambda e^{-R^\alpha Q^\alpha}\right]$ & 5\\
    $H_3$ & Simplified $\tau$-model & $\displaystyle \gamma[1+\varepsilon Q]\,\left[1 + \lambda e^{-R^{2\alpha} Q^{2\alpha}}
      \cos[\tan(\alpha\pi/2)\, R^{2\alpha}Q^{2\alpha}]\right]$ & 5\\
    $H_4$ & 1st-order L\'evy polynomial & $\displaystyle \gamma\,\left[1 + \lambda e^{-R^\alpha Q^\alpha}[1+c_1 L_1(Q|\alpha,R)] \right]$ & 5 \\
    $H_5$ & 3rd-order L\'evy polynomial & $\displaystyle \gamma\,\left[1 + \lambda e^{-R^\alpha Q^\alpha}[1+c_1 L_1(Q|\alpha,R)+ c_3 L_3(Q|\alpha,R)] \right]$ & 6 \\
    $H_6$ & 1st-order derivative & $\displaystyle \gamma\,\left[1 + \lambda e^{-R^\alpha Q^\alpha} + c_1\tfrac{d}{dQ} e^{-R^\alpha Q^\alpha} \right]$ & 5 \\
    $H_7$ & 3rd-order derivative & $\displaystyle \gamma\,\left[1 + \lambda e^{-R^\alpha Q^\alpha} + c_1\tfrac{d}{dQ} e^{-R^\alpha Q^\alpha}+ c_3\tfrac{d^3}{dQ^3} e^{-R^\alpha Q^\alpha} \right]$ & 6 \\[6pt]
    \hline 
  \end{tabular} 
  \ \\[8pt]
  {\small \textbf{Table 1:} Summary of parametrisations tested}
\end{center}
\renewcommand{\arraystretch}{1}
\begin{figure}[htp]
  \begin{center}
    \includegraphics[width=65mm]{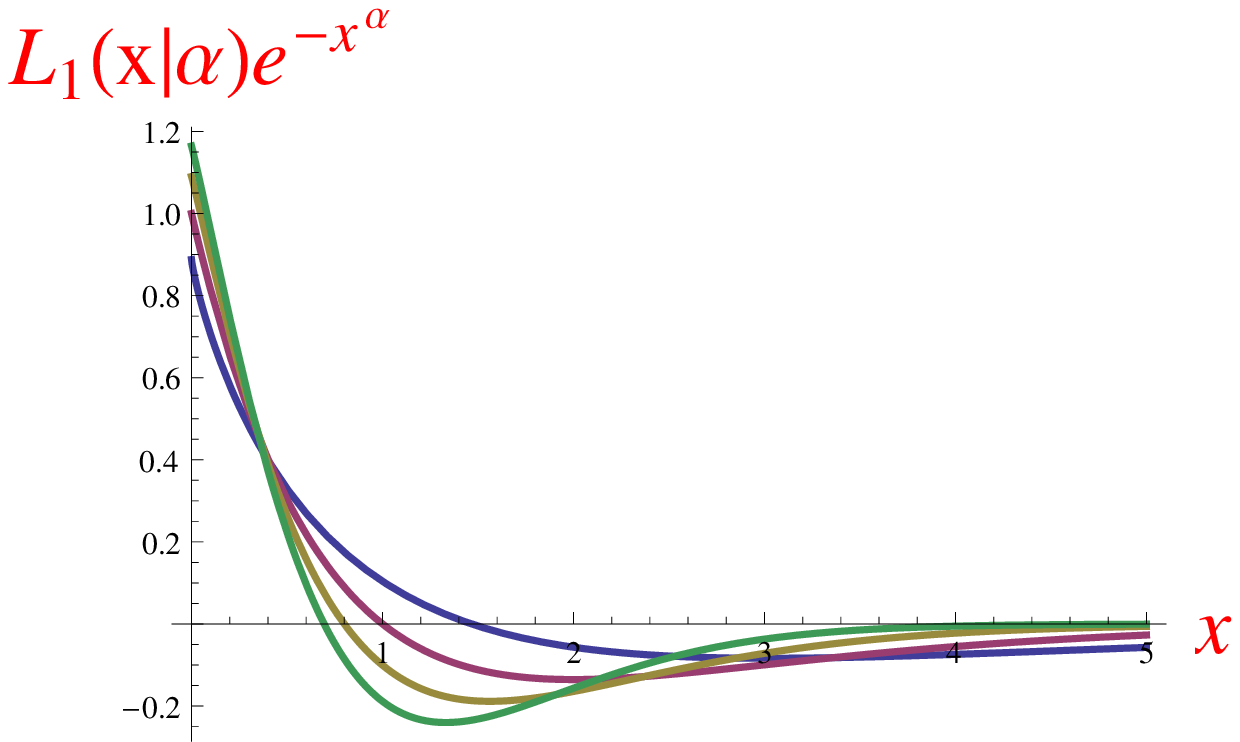}
    \hspace*{4ex}
    \includegraphics[width=65mm]{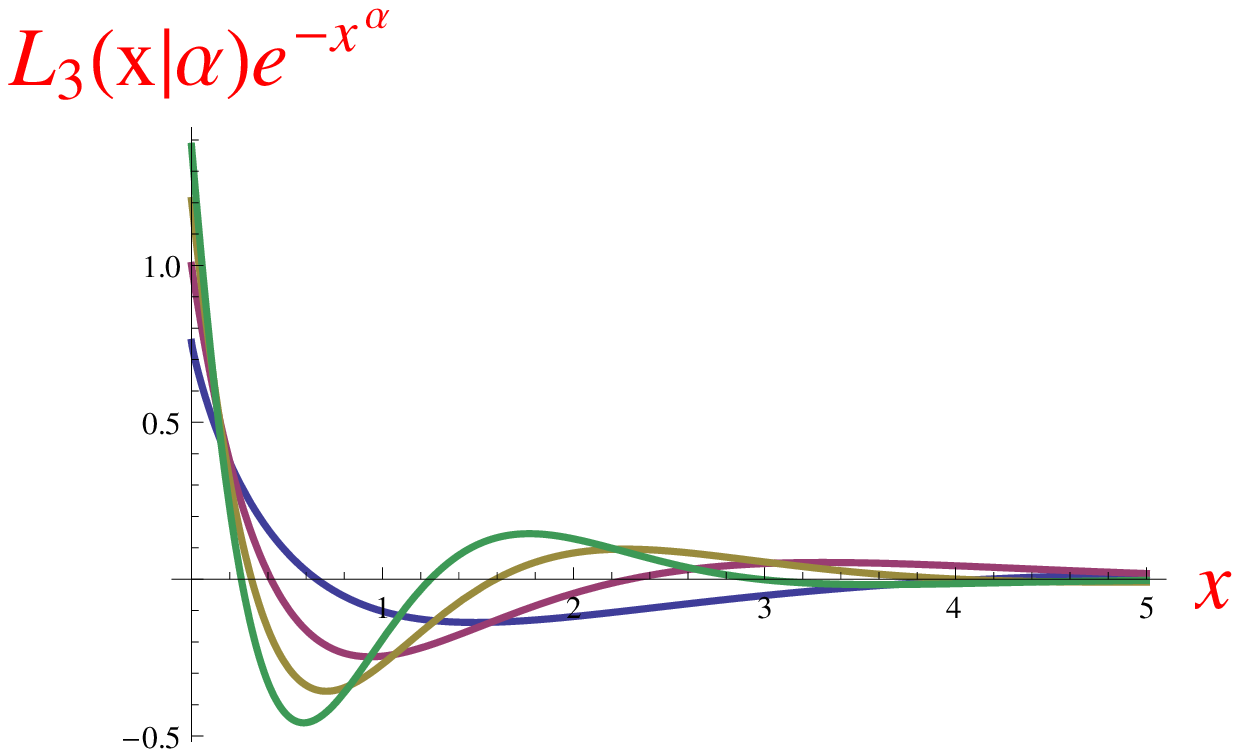}
    \caption{L\'evy polynomials of first and third order times the
      weight function $e^{-x^\alpha}$ for $\alpha = 0.8, 1.0,
      1.2, 1.4$.}
  \end{center}
\end{figure}

\section{Application to L3 binned data}

\noindent
In Table 2, we show the results of applying the Laplace approximation
(\ref{bqh}) to the L3 two-jet data, which is provided in terms of 100
binned values for the correlation function $C(Q_b)$ together with
standard errors $\sigma(C(Q_b))$ in the range $0 < Q < 4$
GeV. Throughout, we used a Gaussian prior $p(\bmtheta_m^*\,|\,H_m)$
with a width which was determined by numerical integration over one of
the L3 data points. To illustrate the contributions of the likelihood,
prior and determinant factors entering $h_m$ in (\ref{bqj}), we have
listed their logarithmic contributions separately in the three columns
headed L, P and F. These quantities are therefore the building blocks
for calculating the odds between any two competing hypotheses.  Thus
one can, for example, deduce that the odds for $H_7$ compared to $H_6$
are $2^{100.6-97.0} \simeq 12{\,:}1$. Also included in Table 2 are the
traditional $\chi^2$ measure (C) and its associated confidence level
(CL).
\begin{center}
  \begin{tabular}{|r@{\hspace*{1.3ex}}l| c| *{6}{r@{.}l|}}
    \hline
    \rule[-3mm]{0mm}{8mm} 
    & Hypothesis & $N_m$
    & \multicolumn{2}{|c|}{L} & \multicolumn{2}{|c|}{P} & \multicolumn{2}{|c|}{F} & \multicolumn{2}{|c|}{$h_m$} 
    & \multicolumn{2}{|c|}{C} & \multicolumn{2}{|c|}{CL} \\ \hline\hline
    $H_1$ & Gauss                       &4& 177&8 & -3&6 & 32&2 & 206&5 &  2&57 &  3&4$\times 10^{-13}$\%\\ \hline
    $H_2$ & Stretched Exponential       &5& 138&5 & -0&5 & 34&0 & 172&0 &  2&02 &  1&5$\times10^{-6}$\%\\ \hline
    $H_3$ & Simplified $\tau$-model     &5&  68&2 & -3&4 & 37&0 & 101&8 &  1&00 &  49&1\%\\   \hline
    $H_4$ & 1st-order L\'evy polynomial &5&  66&2 &  2&2 & 30&3 &  98&8 &  0&97 &  57&3\% \\ \hline
    $H_5$ & 3rd-order L\'evy polynomial &6&  65&9 &  3&8 & 41&6 & 111&3 &  0&97 &  55&7\% \\ \hline
    $H_6$ & 1st-order derivative        &5&  67&3 &  4&2 & 29&1 & 100&6 &  0&98 &  53&0\% \\ \hline
    $H_7$ & 3rd-order derivative        &6&  60&4 &  4&9 & 31&7 &  97&0 &  0&89 &  77&0\% \\ \hline
    \hline
  \end{tabular}
  \ \\[12pt]
  \begin{minipage}{120mm}
    {\small \textbf{Table 2:} Results of fitting parametrisations listed in Table 1.} \\[0pt]
    \hspace*{6ex}
    \begin{tabular}{lll}
      \small
      Legend: 
      & L $\equiv -\lg P(\bD\,|\,\bmtheta_m^*,H_m) \equiv \chi^2/(2\ln 2)$ 
      & $h_m \equiv L + P + F$ \\[2pt]
      & P $\equiv -\lg P(\bmtheta_m^*\,|\,H_m)$
      & C $\equiv \chi^2/(B-N_m)$ \\[2pt]
      & F $\equiv -\lg \sqrt{(2\pi)^{N_m}\det \bmA}$ 
      & CL $\equiv$ \small confidence level \\
    \end{tabular}
  \end{minipage}
\end{center}
It is inappropriate to generalise conclusions based on one specific
dataset with its specific circumstances. The fact that in the two-jet
L3 data the correlation function $C_2(Q)$ drops well below 1.0 for
$0.5 < Q < 2$ GeV, for example, is probably the dominant influence on
the goodness of fit. Under this caveat, we make the following
observations regarding the results shown in Table 2:
\begin{enumerate}

\item At first sight, the Bayes factor and the $\chi^2$ methodologies
  deliver judgements which are rather similar: $H_7$ is consistently
  ranked best, while $H_1$ and $H_2$ are ranked worst (least
  likely). The two methodologies yield vastly different numbers
  when one hypothesis is bad. As shown below, there are surprising
  variations even among the better ones.

\item The determinant plays an important role. For example, factor
  F$=41.6$ for $H_5$ is significantly larger than that of similar
  models $H_4$ and $H_6$ even though the three log likelihoods are
  similar. This can be traced to the fact that the uncertainty in the
  parameters for $H_5$ is larger, as expressed in the width of its
  Gaussian (\ref{bqf}). While $\chi^2$, based only on the likelihood,
  can hardly distinguish between $H_4$ and $H_5$, the contribution of
  the large $H_5$ determinant ensures that the Bayesian odds for $H_4$
  versus $H_5$ are 5800:1. In other words, by taking into account not
  only the best parameter values $\bmtheta_5^*$ but also their
  uncertainties, the Bayes factor could distinguish what $\chi^2$
  could not.

\item Our Bayes factor calculation takes the experimental standard
  errors $\sigma(C(Q_b))$ into account by using (\ref{cmf}) in the
  exponent of the likelihood; in other words, we assume that they are
  Gaussian. We can improve on this approximation by doing a more
  complete Bayesian analysis using not the binned data but the pair
  momenta $\{Q_i\}$ themselves.

\item As Fig.~1 shows, the L\'evy polynomials introduced here are well
  suited to describe one-sided strongly-peaked data. It may be helpful
  to use them, as we have done here, merely as part of
  parametrisations of data to which they show some resemblance.  More
  systematic use in Gram-Charlier or other expansions will be faced
  with issues inherent in all asymptotic series
  \cite{MdK09a,Eggers:2010pm}.

\end{enumerate}

\section{Conclusions}

\begin{enumerate}

\item In hypotheses $H_4$ to $H_7$, we have presented new techniques
  to study deviations from a stretched exponential or
  Fourier-transformed L\'evy shape. Details will be published elsewhere.%

\item The standard measures of fit quality like $\chi^2$ or CL are
  useful in rejecting models which are inconsistent with a given
  dataset. Where two or more models are consistent with the data,
  however, they are unable to select the more probable. The Bayes
  factor (\ref{bqk}) permits quantification of the evidence (relative
  probability) for the validity of models.

\item Besides the likelihood, the prior and determinant also play a
  role, sometimes decisively so.

\item The Laplace approximation (\ref{bqf}) is usually fairly
  accurate, but the assumption of Gaussian errors for count data
  (\ref{cme}), which is made by truncation of the Taylor expansion in
  the data, is of dubious quality.

\item By integrating over parameter space, Bayesian evidence takes
  into account \textit{all} possible values of the parameters, while
  $\chi^2$ and Maximum Likelihood do not.

\item Bayes factors depend linearly on the two priors. This is good in
  that they are made explicit, but bad in the sense that results can
  and do change depending on the choice of priors. 

\item The omission of priors in $\chi^2$ is to its disadvantage as it
  discards important information.

\item It may appear that $\chi^2$ does not need any alternative
  hypothesis to be of use. This is not so, however: the alternative
  implicit in $\chi^2$ is the ``Bernoulli class'' of multinomials
  \cite{Jay03a}. 

\end{enumerate}

\noindent
\textbf{Acknowledgements:} We thank the L3 collaboration for making
its results available electronically \cite{L311a} and the organizers
of WPCF 2011 for support and an excellent atmosphere. This work was
supported in part by the South African National Research Foundation
and by the Hungarian OTKA grant NK--101438.



\end{document}